\documentclass[preprint]{aastex}

\slugcomment{Accepted for publication in The Astrophysical Journal}

\shorttitle{Limits on Birefringence from Radio Galaxies}
\shortauthors{di Serego Alighieri et al.}

\begin{document}


\def\la{\langle }
\def\ra{ \rangle }
\def\be{\begin{equation}}
\def\ee{\end{equation}}
\def\bea{\begin{eqnarray}}
\def\eea{\end{eqnarray}}
\def\gmf{\gamma_{5}}
\def\tQ{\tilde Q}
\def\edth{\;\raise1.0pt\hbox{$'$}\hskip-6pt\partial\;}
\def\baredth{\;\overline{\raise1.0pt\hbox{$'$}\hskip-6pt
\partial}\;}

\title{Limits on Cosmological Birefringence from the Ultraviolet Polarization of
Distant Radio Galaxies}
\author{Sperello di Serego Alighieri}
\affil{INAF - Osservatorio Astrofisico di Arcetri,
Largo E. Fermi 5, I-50125 Firenze - Italy}

\author{Fabio Finelli\altaffilmark{1} and Matteo Galaverni}
\affil{INAF-IASF Bologna, 
Via Gobetti 101, I-40129 Bologna - 
Italy}
\altaffiltext{1}{INFN, Sezione di Bologna,
Via Irnerio 46, I-40126 Bologna - Italy}

\begin{abstract} 
We report on an update of the test on the rotation of the plane of linear polarization for light
traveling over cosmological distances, using a comparison between the measured direction of the 
UV polarization in eight RG at $z>2$ and the direction predicted by the model of scattering of
anisotropic nuclear radiation, which explains the polarization. No rotation is detected within a few
degrees for each galaxy and, if the rotation does not depend on direction, then the all--sky--average
rotation is constrained to be $\theta = -0.8^o\pm2.2^o$.
We discuss the relevance of this result for constraining cosmological birefringence, when this is  
caused by the interaction with a cosmological pseudo-scalar field or by the presence of a Cherns-Simons term.

\end{abstract}

\maketitle

\section{Introduction}

The possibility that the propagation of light through our universe might suffer from chiral effects,
which could rotate the plane of polarization, arises in a variety of important contexts, such as the presence of 
a cosmological pseudo-scalar condensate, Lorentz invariance violation and charge parity and time (CPT) violation, neutrino number asymmetry,  
and the Einstein equivalence principle (EEP) violation (see \citet{Nio07} for a review).
The simplest form for modeling cosmological birefringence - a frequency independent rotation of the plane of linear polarization -
is described by the interaction of a pseudo-scalar field $\phi$ with photons through a term \citep{Kol90,Raf96}:
\begin{equation}
\label{eq:1}
\mathcal{L}_{int}=-\frac{g_\phi}{4}\phi  F_{\mu\nu}\tilde{F}^{\mu\nu}\,,
\end{equation}
where $g_\phi$ is the coupling constant,
$F^{\mu\nu}$ is the electromagnetic tensor and
$\tilde{F}^{\mu\nu}\equiv\frac{1}{2}\epsilon^{\mu\nu\rho\sigma}
F_{\rho\sigma}$ its dual. $\phi$ could 
be a fundamental field or an effective description for cosmological birefringence due to Lorentz violation \citep{Car89}. 

Indeed several efforts have been devoted to look for evidence of rotation of the plane of polarization:
since we expect tiny effects on the basis of laboratory experiments, cosmological distances 
are required to have measurable effects and therefore the obvious approach has been to look
for rotation in the most distant sources in the universe. What is required for this test is then a polarized
distant source, for which the polarization orientation can be predicted: the predicted orientation is then
compared with the measured one, looking for a rotation between the two. Radio galaxies (RG) are very good candidates, 
since these astrophysical objects are often polarized, both at radio and at UV-optical wavelengths, 
and are found at very high redshifts \citep{Mil08}. Since the first successful detection of anisotropies 
in polarization of the cosmic microwave background (CMB) by DASI in 2002 \citep{dasi}, also the 
CMB polarization pattern has become an important 
test for cosmological birefringence, which could probe the propagation of light back to 
the recombination surface, i.e. up to a redshift as high as $z \sim 1100$.

Cosmological birefringence was first constrained from RG observations, since these 
were the first cosmological sources providing information on polarization.
\citet{Car89} have used the fact that the 
distribution of the difference between the position angle (P.A.) of the radio axis and the 
P.A. of the E vector of linear radio polarization in distant RG
($0.4<z<1.5$) peaks around $90^o$ to argue that this phenomenon is intrinsic to the source and 
therefore to put limits ($|\theta| \le 6.0^o$ at the 95\% confidence 
level) on the rotation of the plane of polarization for radiation traveling over
cosmic distances. Later \citet{Cim93} used the perpendicularity between the 
optical/UV axis and the linear optical/UV polarization of distant RG ---  this perpendicularity is
expected since the polarization and the elongation are due to scattering of anisotropic nuclear 
radiation --- to show that the plane of polarization is not rotated by more than $10^o$ for every
distant RG with a polarization measurement up to $z=2.63$. The advantage of
the test using the optical/UV polarization over that using the radio one is that
it is based on a physical prediction of the orientation of the polarization due to
scattering, which is lacking in the radio case, and that it does not require a correction for
the Faraday rotation, which is considerable in the radio but negligible in the optical/UV.

A few years later \citet{Nod97} claimed to have found a rotation, independent of the
Faraday one, in the radio polarization of distant RG. However several authors
\citep{War97,Eis97,Car97,Lor97} have independently and convincingly 
argued against this claim, and additional unpublished data \citep{Lea97} on the lack
of rotation for the radio polarization of distant RG have been reported \citep{Car98}.

As already said, the observed polarization of the CMB has recently been used to put stringent
constraints on cosmological birefringence, 
which would modify the linear polarization pattern created first by Thomson scattering and then by 
reionization, and generate correlations between time and magnetic field and between electric and magnetic fields, 
otherwise zero in a standard cosmological scenario. 
By using the constant angle approximation - we denote $\bar \theta$ the rotation angle in the following - 
for the integrated rotation of the linear polarization plane along the line of sight 
\citep{Lue98}, the observed power spectra are proportional to the power spectra on the last scattering surface through 
trigonometric functions of $\bar \theta$. Several constraints, summarized in  Table 1, have been obtained within this approximation 
(see however \citet{Fin08} for the limits of the constant angle approximation).

In this paper we report on an update of 
the test using the UV polarization of distant RG, because several new polarization
information has become available on very distant RG since this test was last performed 
\citep{Cim93}, and discuss its implications in various contexts. 
Our paper is organized as follows: after this introduction, we describe the set of observations on UV polarization 
and the constraints on the rotation angle. We then discuss the implications of this constraints for cosmological 
birefringence caused by a pseudo-scalar field (playing the role of dark matter or dark energy) and by a Cherns-Simons term 
respectively in Sections 3 and 4. In Section 5 we conclude.

\section{Limits on the rotation of UV linear polarization of radio galaxies at $z>2$}

The birefringence test based on the UV polarization of RG is independent, complementary
and placed at a different frequency with respect to those based on the radio polarization of distant RG
and on the CMB polarization.
The UV polarization test has also some advantages over the other tests.
The main advantage over the test based on the radio polarization is that the UV and the CMB tests are based on
a clear prediction of the polarization angle, given by the scattering physics, while a clear prediction is
lacking for the radio polarization angle, which is only phenomenologically found to peak at about $90^o$ and $0^o$ 
from the radio axis, without a clear understanding of the physics behind it \citep{Cla80}.
Distant RG observations provide a snapshot integrated up to a much smaller redshifts ($z \simeq {\rm few}$) with 
respect to the CMB one: as it occurs for CMB and SN Ia in probing the expansion history, the combination of CMB 
and RG may be very useful to constrain the cosmological birefringence.
Being based at short wavelengths, the UV test is practically immune from Faraday rotation by intervening magnetic fields 
along the line of sight, which instead is relevant for radio and - to a smaller extent - for microwave observations
\citep{Sca97}, reminding however that the Faraday rotation can be corrected for, since it depends on
frequency, while birefringence does not.

After the first birefringence test based on the UV polarization of distant RG by \citet{Cim93},
the test has been repeated by other authors. In particular, the RG 3C 265 at z=0.811 is a suitable source,
because its misalignment between the radio and optical/UV axes provides a crucial check of the scattering hypothesis
\citep{diS96} and because its bright extensions allow to build up a good polarization map \citep{Tra98}, in which 
the perpendicularity of the polarization vectors can be tested for each of the several tens of independent measurements
at different locations.
Indeed the spectacular polarization pattern of 3C265 has been used by Wardle et al. (1997) to rule out the birefringence
claimed by Nodland \& Ralston (1997). Since then, several new polarization measurements for distant RG have
become available and an update of the birefringence test has become desirable, in particular using the most distant
ones, as a complement of the similar test performed using the CMB polarization.

In order to perform the best test now possible with RG, we have selected all RG with
$z>2.0$, with the degree of linear polarization $P$ larger than 5\% in the far UV 
(at $\sim$ 1300 \AA, rest frame), and with elongated optical morphology at these wavelengths, since
these are the marking characteristics of the presence of scattered nuclear radiation \citep{diS94},
and can therefore lead to a safe test of the polarization rotation \citep{diS95}. 
The relevant data are collected from the literature in Table 2. The second-last column of the 
table lists the difference between the P.A. of the linear UV polarization and the P.A. of the UV axis, 
which we have measured on the available images in the rest-frame UV, and is shown in Figure 1.
According to the scattering model, these two directions should be perpendicular for every object
in our sample.
The fact that the P.A. difference is close to $90^o$ for every object, actually compatible with 
$90^o$ within the accuracy of the measurements, 
puts stringent constraints on any possible rotation $\theta$ of the polarization plane for light traveling
to us from each RG, as listed in the last column of the table. Assuming that the rotation
of the polarization plane should be the same in every direction (as is done in the CMB case), we can
set the average constraint $\theta = -0.8^o\pm 2.2^o$, as listed in the last row of the table.

\section{Constraint on Cosmological Pseudo-Scalar Fields}

Upper limits on the linear polarization rotation angle $\theta$
can be used to constrain cosmological birefringence 
caused by the coupling of the electromagnetic field to pseudo-scalar fields, suggested to solve the strong 
charge and parity (CP) problem \citep{Pec77}. The existence of light pseudo-scalar particles \citep{Wei77} 
is very relevant in cosmology, since these are 
viable candidate either for dark matter \citep{Kol90} 
or for dark energy \citep{Fri95}, depending on their (effective) mass. A pseudo-scalar field $\phi$ is predicted to be coupled 
to photons as can be read from the Lagrangian of the electromagnetic-$\phi$ sector:
\be
\mathcal{L}= -\frac{1}{4}F_{\mu\nu}F^{\mu\nu}-\frac{1}{2}\nabla_{\mu}\phi\nabla^{\mu}\phi - V(\phi)
-\frac{g_{\phi}}{4}\phi F_{\mu\nu}\tilde{F}^{\mu\nu}
\label{total}
\ee
where $V(\phi)$ is the potential for the pseudo-scalar field. 
At the lowest order in fluctuations, the photon is coupled to the time 
derivative of the cosmological value of $\phi$, which is governed by the potential. Different time evolutions of $\phi$ 
lead to different values for the resulting cosmological birefringence, and therefore in the following 
two subsections we consider representative 
cosmological scenarios involving totally different values for the time variation of $\phi$ 
\footnote{Note that the CMB polarization auto and cross spectra depend on the time variation 
of $\phi$ and in many cases the constant angle approximation is a poor description of cosmological 
birefringence in CMB anisotropies \citep{Fin08}.}. 

\subsection{Dark Matter Pseudo-scalar Field}
\label{Sect:DM}

We consider as potential in Equation (\ref{total}):
\be
V(\phi)=m^2 f_a^2 \left(1-\cos\frac{\phi}{f_a}\right)
\label{potential_DM}
\ee
where $m$ is the mass and $f_a$ is the energy scale at which the Peccei-Quinn symmetry is broken.
In the dark matter regime the pseudo-scalar
field oscillates near the minimum of the potential,
therefore $V(\phi)\simeq m^2\phi^2/2$.
The evolution of the field as a function of cosmic time $t$ is \citep{Fin08}
\bea
\phi(t)
&\simeq&\sqrt{\frac{3\Omega_\mathrm{ MAT}}{\pi}}\frac{H_0 M_\mathrm{ pl}}{2 m a^{3/2}(t)}\nonumber\\
\label{E:phi_t_1}
& &
\sin\left[m t \sqrt{1-\left(1-\Omega_\mathrm{ MAT}\right)\left(\frac{3H_0}{2m}\right)^2}\right]\,.
\label{phi:EVOL}
\eea
where $\Omega_\mathrm{ MAT}$ is the density parameter for $\phi$ nowadays (which we consider equal to the dark matter one),
$H_0$ is the Hubble constant, $M_\mathrm{ pl}$
is the Planck mass. Averaging through the oscillations, the evolution of the scale factor is given by \citep{Fin08}:
\begin{eqnarray}
\label{a:LCDM}
a(t) & \simeq &\left(\frac{\Omega_\mathrm{ MAT}}{1-\Omega_\mathrm{ MAT}}\right)^\frac{1}{3}\nonumber\\
& &\left[\sinh \left(\frac{3}{2}\sqrt{1-\Omega_\mathrm{ MAT}} H_0 t\right)\right]^\frac{2}{3}\,.
\end{eqnarray}

Considering photon propagation in a homogeneous pseudo-scalar background ($\phi=\phi(\eta)$) 
the Fourier transform of the electromagnetic vector potential in the basis of left and right circular polarized modes 
in the plane transverse 
to the direction of propagation in the Coulomb gauge ($\nabla\cdot{\bf A}=0$) is:
\be
\label{eq:Apm}
\tilde{A}_{\pm}^{\prime\prime}(k,\eta)+\left[ k^2 \pm g_{\phi} \phi^{\prime}  k \right] \tilde{A}_{\pm}(k,\eta)=0\,,
\ee
where $\prime$ denotes derivative respect to conformal time $\eta$ ($d\eta=dt/a(t)$, \citet{Fin08}).
The linear polarization rotation angle is given by:
\begin{eqnarray}
\theta_{\rm DM}(z)&=&\frac{g_\phi}{2}\left[\phi\left(\eta_0\right)-\phi\left(\eta\right)\right]\nonumber\\
&=&\frac{1}{4}\sqrt{\frac{3\Omega_\mathrm{ MAT}}{\pi}}\frac{g_\phi M_\mathrm{ pl}H_0}{m}
\left(\frac{1}{a_0^{3/2}}-\frac{1}{a^{3/2}}\right)\nonumber\\
&=&- \frac{1}{4}\sqrt{\frac{3\Omega_\mathrm{ MAT}}{\pi}}\frac{g_\phi M_\mathrm{ pl}H_0}{m } \left[1-(1+z)^{3/2}\right]\,.
\end{eqnarray}
Fixed the average redshift ($\bar{z}=3$),
$H_0=72\, \mathrm{km \, s}^{-1} \, \mathrm{Mpc}^{-1}$,
$M_\mathrm{ pl}\simeq 1.22\times10^{19}$ GeV,
and $\Delta\theta<5.0^o$ we obtain a constraint in the
plane $(\log_{10} m\,\left[\mbox{eV}\right],\, \log_{10} g_\phi\, \left[\mbox{eV}^{-1}\right])$, as from 
Figure~\ref{mg01_rg}, which we superimpose with the one obtained in \citet{Fin08}.

\subsection{Dark Energy pseudo-scalar field}
\label{Sect:DE}

An ultralight pseudo Nambu-Goldstone boson could drive an accelerated expansion of the universe, as proposed by 
\citet{Fri95}, by considering a simple shift of the potential 
in Equation (\ref{potential_DM}): 
\be
V(\phi) = M^4 \left[1+\cos(\phi/f)\right]
\label{potential_DE}
\ee
with $M$ and $f$ mass and energy scale for the dark energy 
case, respectively (note that these numbers and $g_\phi$ may be quite different from the dark matter case). When $\phi$ acts as dark energy, 
it is presently rolling toward the bottom of the potential (located at $\phi = \pi f$) with small velocity: 
in the future, $\phi$ will roll around the bottom of the potential and will be another matter component added to cold dark
matter (CDM).

The linear polarization angle $\theta$ is related to the variation of $\phi(\eta)$:
\be
\label{theta:DE}
\theta(\eta)=\frac{g_\phi}{2} \left[\phi(\eta_0)-\phi(\eta) \right]\,.
\ee
and the evolution of $\phi$ is determined solving the following system of equations:
\be
\left\{
\begin{array}{l}
\ddot{\phi}+3H\dot{\phi}-\frac{M^4}{f}\sin\frac{\phi}{f}=0\,,\\
H^2=\frac{8\pi}{3 M_\mathrm{ pl}^2}\left(\rho_\mathrm{ RAD}+\rho_\mathrm{ MAT}+\rho_{\phi}\right)\,.
\end{array}
\right.
\ee
We solve it numerically fixed $M=8.5\times10^{-4}$ eV, $f=0.3 M_\mathrm{ pl}/\sqrt{8\pi}$,
$\phi_i/f=0.25$ and $\dot{\phi}_i=0$ \citep{Abr08}: see Figure~\ref{plot::modelPNGBb} for the evolution of the 
critical densities for matter ($\Omega_\mathrm{ MAT}$),
dark energy ($\Omega_{\phi}$) and for the parameter $w_\phi\equiv p_\phi/\rho_\phi$ of the dark energy equation of state.

Figure~\ref{plot::ThetaPNGB2b} shows the variation of $\phi/f$ as a function of $\ln\,a/a_0$.
In the region probed by high-redshift RG ($\bar{z}=3$) there is a variation
of the pseudo-scalar field the order $\phi/f\sim1.1$. Therefore Equation~(\ref{theta:DE}) can be used to obtain
an upper limit on $g_\phi$:
\begin{eqnarray}
& &-5.0<\theta<3.4\nonumber\\
& &\Longrightarrow -2.2 \times 10^{-28} {\rm eV}^{-1} < g_\phi < 1.5 \times 10^{-28} {\rm eV}^{-1}
\end{eqnarray}

Let us also consider a runaway potential like:
\be
V(\phi) = V_0 \exp \left( - \lambda \sqrt{8 \pi} \frac{\phi}{M_{\rm pl}} \right) \,.
\label{expotential_DE}
\ee
The above potential has only $M_{\rm pl}$ as physical scale, differently from the one in Equation (8).
The resulting dark energy model is stable for $\lambda < \sqrt{2}$ and has an equation of state $p_\phi=w_\phi \rho_\phi$, with 
$w_\phi = -1 + \lambda^2/3$, constant in time \citep{CLW}. The evolution of the scale factor in this cosmological model 
can therefore be found analytically \citep{GF}, as also the evolution of the scalar field. We therefore 
give the analytical formula for the rotation angle:
\begin{eqnarray}
\theta_{\rm DE}(z) &=& \frac{g_\phi}{2}\left[\phi\left(\eta_0\right)-\phi\left(\eta\right)\right]\nonumber\\
&=& g_\phi M_{\rm pl} \sqrt{\frac{1+w_\phi}{3}} \frac{1}{-w_\phi} \left[ \mbox{arcsinh} 
\left( \sqrt{\frac{\Omega_\phi}{1-\Omega_\phi}} \right)  \right. \nonumber \\ & & \left. 
- \mbox{arcsinh} \left( \sqrt{\frac{\Omega_\phi}{1-\Omega_\phi}} a^\frac{- 3 w_\phi}{2}   \right) \right] \,,
\end{eqnarray}
where $\Omega_\phi$ is the dark energy fraction at present time. Figure 5 shows the value of 
$\theta_{\rm DE}(z=2.80)$ as a function of $(\Omega_\phi, w_\phi )$. By considering 
$\theta_{\rm DE}(z=2.80) \simeq 0.2 g_\phi M_{\rm pl}$ as an representative value, 
we obtain $|g_\phi | \lesssim {\rm few} \times {\cal O} (10^{-29}) {\rm eV}^{-1}$.

\section{Constraints on Chern-Simons Theory}
\label{Sect:CS}
We consider the following Lagrangian:
\be
\mathcal{L}=-\frac{1}{4}F_{\mu\nu}F^{\mu\nu}
-\frac{1}{2}p_\mu A_\nu \tilde{F}^{\mu\nu}
\ee
where $p_\mu=(p_0,\mathbf p)$ is a constant 4-vector and $A_\nu$ the vector potential \citep{Car89}.

The corresponding dispersion relation for an electromagnetic-wave $k^\mu=\left(\omega,\mathbf{k}\right)$
is \citep{Car89}:
\be
\omega^2-k^2=\pm\left(p_0 k-\omega p \cos\alpha\right)\left[1-\frac{p^2 \sin^2\alpha}{\omega^2-k^2}\right]^{-\frac{1}{2}}
\ee
where $\alpha$ is the angle between $\mathbf p$ and $\mathbf k$.
The angle by which the plane of polarization rotates is half of the difference phase,
since $p_\mu$ is expected to be small, therefore the dispersion relation can be expanded at
first order in $p_\mu$:
\be
k=\omega\mp \frac{1}{2} \left(p_0-p\cos \alpha\right)\,.
\ee
For a wave traveling a distance $L$ the linear polarization
vector rotates by:
\be
\theta=-\frac{1}{2}\left(p_0-p \cos\alpha\right)L
\ee
independent of wavelength.

In a $\Lambda$CDM universe the evolution of the scale factor in terms of cosmic time is given by Equation~(\ref{a:LCDM}),
therefore the relation between $t$ and redshift is:
\be
t=\frac{2}{3
H_0\sqrt{1-\Omega_\mathrm{MAT}}}\mbox{arcsinh}\left[\sqrt{\frac{1-\Omega_\mathrm{MAT}}{\Omega_\mathrm{MAT}}}\left(\frac{1}{1
+z}\right)^{\frac{3}{2}}\right]
\ee
Fixed $L=t$ for the distance traveled by  photons, the linear polarization plane from redshift $z$ to nowadays
rotates by:
\bea
\theta&=&-\frac{1}{2}\left(p_0-p \cos\alpha\right) t \nonumber\\
&=&-\frac{p_0-p \cos\alpha}{3 H_0\sqrt{1-\Omega_\mathrm{MAT}}}
\left\{\mbox{arcsinh}\left[\sqrt{\frac{1-\Omega_\mathrm{MAT}}{\Omega_\mathrm{MAT}}}\right]\right.\nonumber\\
& &\left.-
\mbox{arcsinh}\left[\sqrt{\frac{1-\Omega_\mathrm{MAT}}{\Omega_\mathrm{MAT}}}\left(\frac{1}{1+z}\right)^{\frac{3}{2}}\right]
\right\}
\eea

Fixed $\Omega_\mathrm{MAT}=0.3$,
$H_0 = 100 \, h \, \mathrm{ km \, s}^{-1} \, \mathrm{Mpc}^{-1}=2.13\,h\,\times10^{-33}$ eV
and $\bar{z}=3$:
\be
\left|p_0-p \cos\alpha\right|<\theta\times5.2\, h\,\times 10^{-33}\,\mbox{eV}
\ee
If $h=0.72$ and $\theta<5.0^o$:
\be
\left|p_0-p \cos\alpha\right|<3.2\times 10^{-34}\,\mbox{eV} \,,
\ee
which updates the constraint given in \citet{Car89} 
for a matter-dominated universe to one valid for the present cosmological concordance model.

\section{Conclusions}

Every single existing measurement of the UV linear polarization in RG at $z>2$, due to scattering of
anisotropic nuclear radiation, excludes that the polarization plane rotates by more than a few degrees while the 
light travels from the source to us for more than 3/4 of the universe lifetime, confirming previous 
results at lower redshifts \citep{Cim93, War97}.
The all-sky-average constraint derived on the rotation of the polarization from the set of observations considered in this paper 
($\theta = -0.8^o\pm 2.2^o$) is independent, but consistent with the constraints derived from CMB observations. 
We have studied the implications of this constraint on physical models of cosmological birefringence, showing how 
observations at high redshifts as those of RG are complementary to CMB anisotropies, as already occurs for 
SN Ia and CMB in measuring the expansion history. In the framework of theoretical models associating the
cosmological birefringence with the variation of the Newton constant our results increase our confidence in
the validity of the EEP, on which all metric theories of gravity are based. 
An improvement in both quantity and quality of the measurements of the UV linear polarization in RG 
at high redshift should be possible in the future with the coming generation of giant optical telescopes 
\citep{Gil08,Nel08,Joh08}, and would narrow the constraint 
on $\theta$ to a level smaller than what is now possible with RG and CMB.

\section*{Acknowledgements} 
We thank Wei-Tou Ni for his encouragement to publish this work. We also thank Marc Kamionkowski and the two referees for helpful comments.

\clearpage

\begin{figure}
\includegraphics[width=8.6cm]{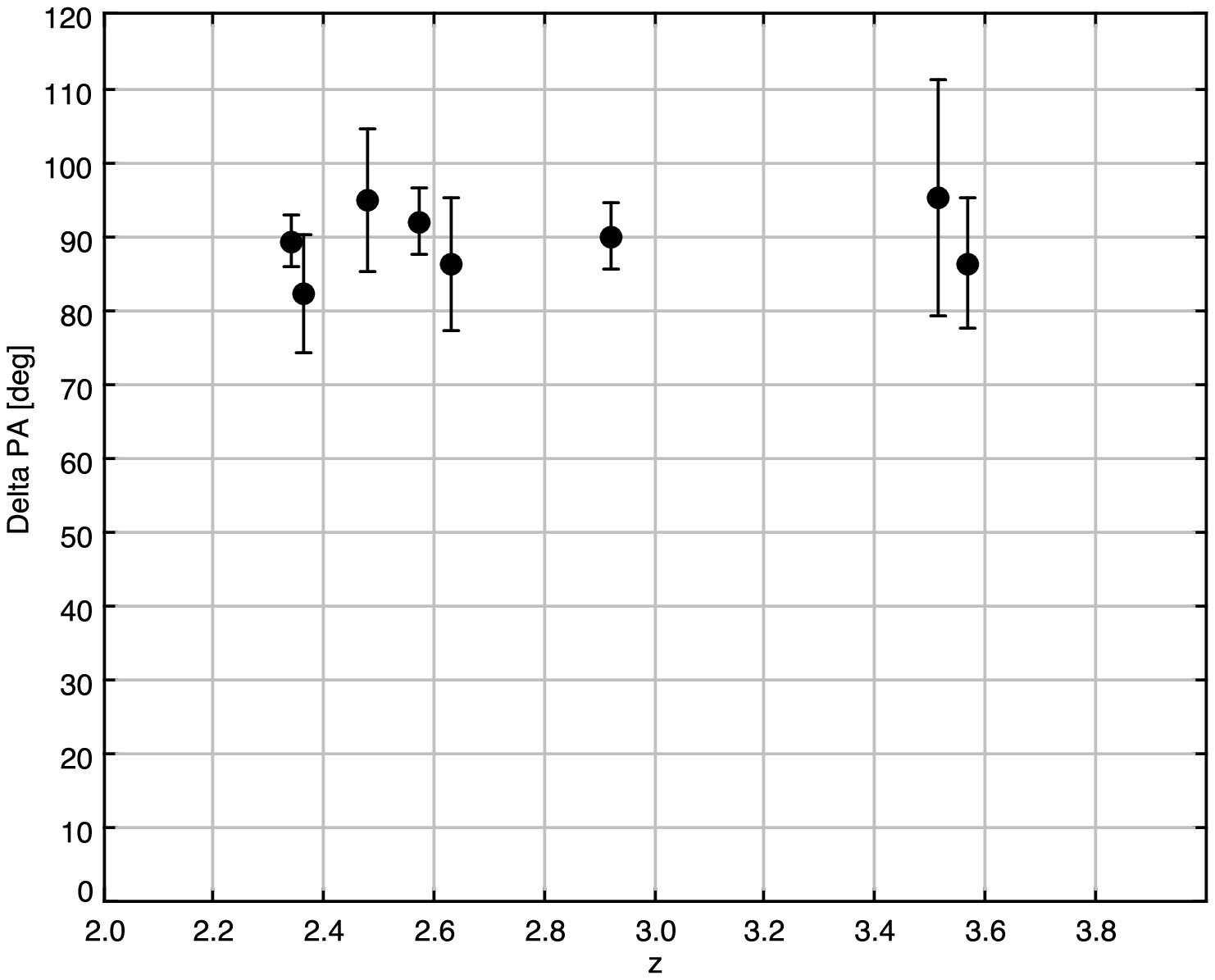}
\caption{Angle between the direction of linear polarization in the UV and the direction of the UV
axis for RG at $z>2$. The angle predicted by the scattering model is $90^o$.}
\label{fig_PA}
\end{figure}

\begin{figure}
\begin{center}
\includegraphics[width=8.6cm]{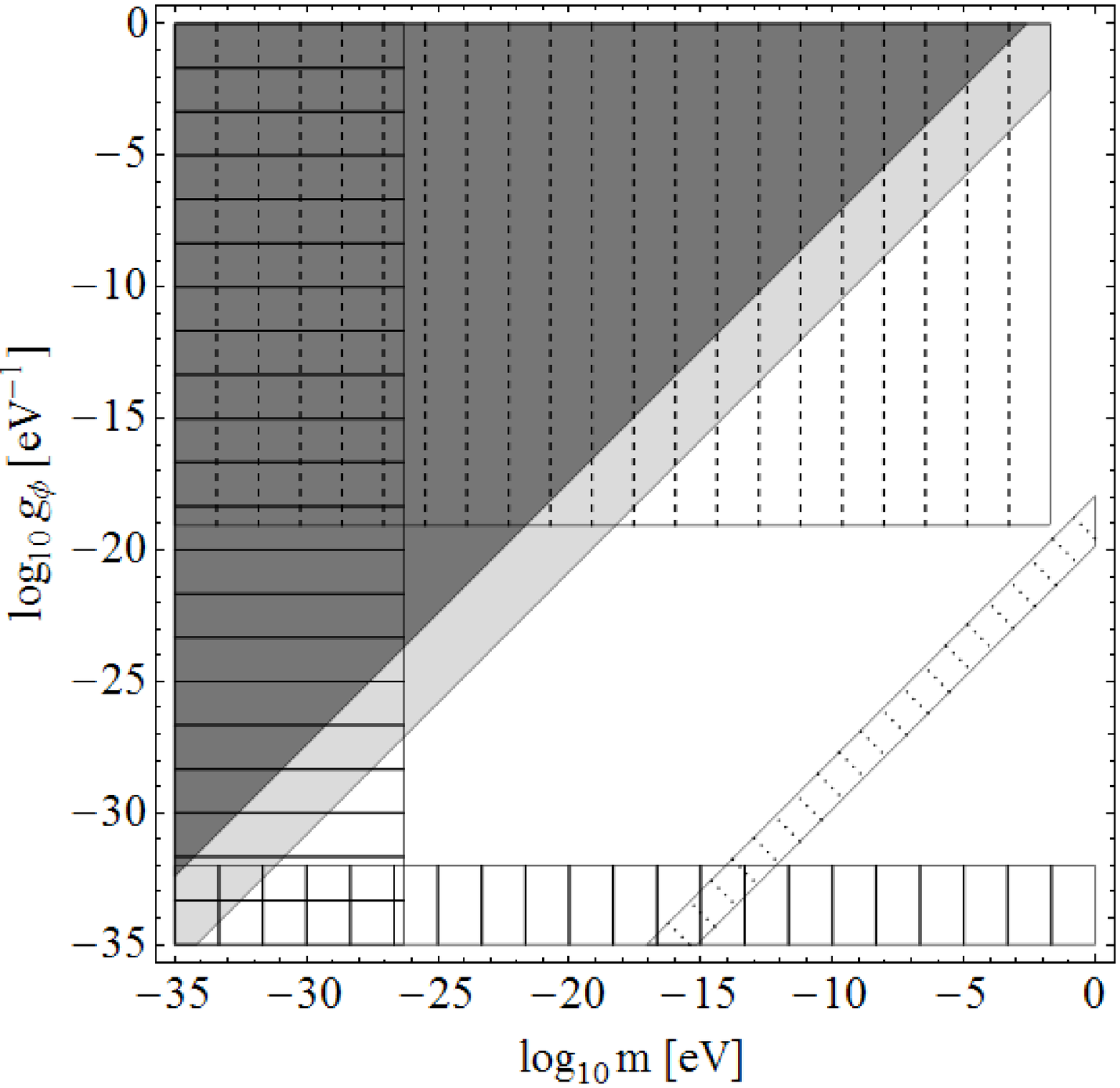}
\caption{Plane $(\log_{10} m\,\left[\mbox{eV}\right],\, \log_{10} g_\phi\, \left[\mbox{eV}^{-1}\right])$:
region excluded by CAST \citep{And07} (white with dashed vertical lines),
region where $\left|\theta_\mathrm{CMB}(\Omega_\mathrm{ MAT}=0.3,m,g_\phi)\right|>10^o$
obtained by the constant angle approximation in \citet{Fin08} (light gray region),
region where $\left|\theta_\mathrm{HzRG}(\Omega_\mathrm{ MAT}=0.3,m,g_\phi)\right|>5.0^o$
(dark gray with dashed vertical lines),
$\left(m,\,g_\phi\right)$ values expected in main QCD
axion models (dotted slanted lines),
region where the mass of the pseudo-scalar field is too small in order
to explain dark matter ($m<3H_\mathrm{ eq}$) (white with horizontal lines),
and
region where PQ symmetry is broken at energies higher than Planck scale
($f_a>M_\mathrm{ pl}$) (white with vertical lines).}
\label{mg01_rg}
\end{center}
\end{figure}

\begin{figure}[htdb]
\begin{center}
\includegraphics[width=8.6cm]{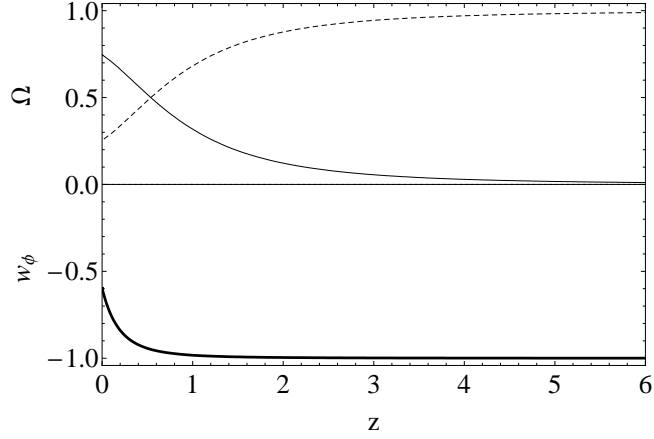}
\caption{Dashed line: $\Omega_\mathrm{ MAT}$,
thin continuous line: $\Omega_{\phi}$, thick continuous line $w_\phi$, in terms of the natural
logarithm of the scale factor (from $\ln\,a\simeq-15$ to nowadays $\ln\,a_0=0$).
Fixed $M=8.5\times10^{-4}$ eV, $f=0.3 M_\mathrm{ pl}/\sqrt{8\pi}$,
$\phi_i/f=0.25$ and $\dot{\phi}_i=0$.
}
\label{plot::modelPNGBb}
\end{center}
\end{figure}

\begin{figure}[htdb]
\begin{center}
\begin{tabular}{c}
\includegraphics[width=8.6cm]{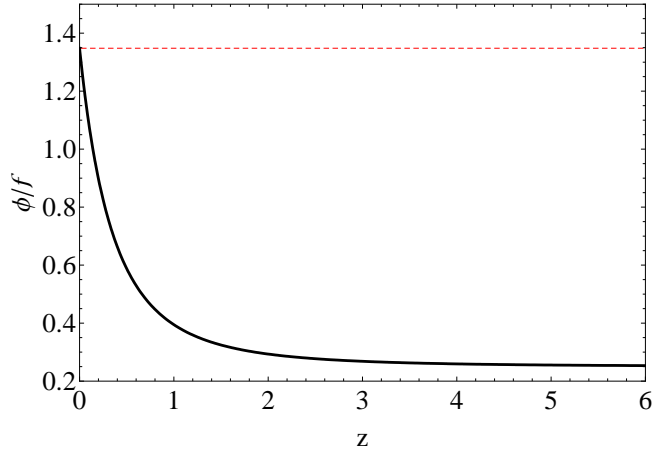}
\end{tabular}
\caption{Evolution of pseudo Nambu--Goldstone boson field $\phi/f$ in terms of the natural
logarithm of the scale factor from $\ln\,a\simeq-15$ to nowadays $\ln\,a_0=0$.
Fixed $M=8.5\times10^{-4}$ eV, $f=0.3 M_\mathrm{ pl}/\sqrt{8\pi}$,
$\phi_i/f=0.25$ and $\dot{\phi}_i=0$.
}
\label{plot::ThetaPNGB2b}
\end{center}
\end{figure}

\begin{figure}[htdb]
\begin{center}
\begin{tabular}{c}
\includegraphics[width=8.6cm]{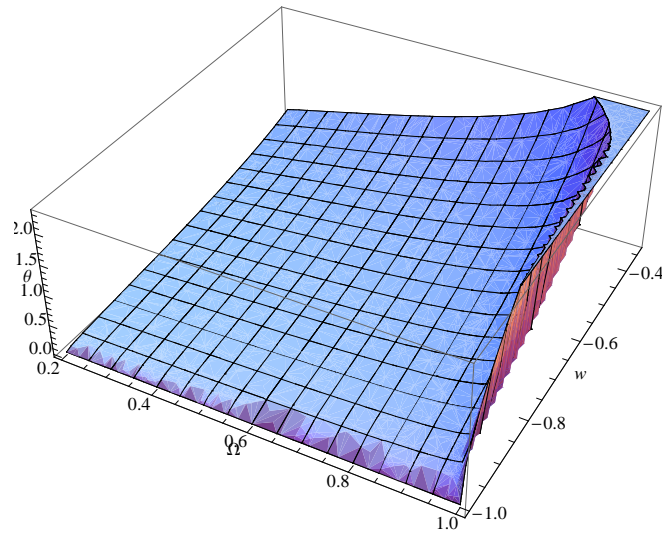}
\end{tabular}
\caption{Three-dimensional plot of $\theta_{\rm DE}(z=2.80)$ as a function of $(\Omega_\phi, w_\phi )$ in the range $\Omega_\phi =
\left[ 0.2, 1 \right]$ and $w_\phi = \left[ -0.34, -1 \right]$.}
\label{plot::Thetaexponential}
\end{center}
\end{figure}

\clearpage

\begin{deluxetable}{lcc}
\tablecaption{Constraints on Linear Polarization Rotation $\bar{\theta}$ in the Constant
Angle Approximation.
}

\tablehead{
\colhead{Data Set} & \colhead{$\bar{\theta}\;(2\sigma)\;\mbox{(deg)}$} & \colhead{Reference}
}
\startdata
WMAP3 and Boomerang (B03) & $-13.7<\bar{\theta}<1.9$ & 1 \\
WMAP3 & $-8.5<\bar{\theta}<3.5$ & 2 \\
WMAP5 & $-5.9<\bar{\theta}<2.4$ & 3 \\
QUaD & $-1.2<\bar{\theta}<3.9$ & 4 \\
WMAP7 & $-5.0<\bar{\theta}<2.8$ & 5 \\
\enddata
\tablerefs{(1) \citet{Fen06}; (2) \citet{Cab07}; (3) \citet{Kom08}; (4) \citet{Wuo08}; (5) \citet{Kom10}.}
\end{deluxetable}

\clearpage

\begin{deluxetable}{lrrlccccc}
\tabletypesize{\scriptsize}
\rotate
\tablecaption{Linear Far UV Scattering Polarization in Distant RG.}
\tablewidth{0pt}
\tablehead{
\colhead{RG Name} & \colhead{RA. (deg)}  & \colhead{Dec. (deg)}  & \colhead{z}    & \colhead{P (\%)}    & \colhead{Pol. P.A. (deg)
} & \colhead{UV P.A. (deg)} & \colhead{$\Delta$P.A. (deg)} & \colhead{$\theta\;(1\sigma)\;\mbox{(deg)}$}
}
\startdata
MRC 0211-122 & 33.5726 & -11.9793 & 2.34 & 19.3$\pm$1.15\tablenotemark{a} & 25.0$\pm$1.8 & 116$\pm$3\tablenotemark{b} & 89.0$\pm
$3.5 & $-4.5<\theta<2.5$\\
4C -00.54    & 213.3131 & -0.3830 & 2.363 & 8.9$\pm$1.1\tablenotemark{c} & 86$\pm$6 & 4$\pm$5\tablenotemark{b} & 82$\pm$8 & $-16
<\theta<0$\\
4C 23.56a    & 316.8111 & 23.5289 & 2.482 & 15.3$\pm$2.0\tablenotemark{c} & 178.6$\pm$3.6 & 84$\pm$9\tablenotemark{d} & 94.6$\pm
$9.7 & $-5.1<\theta<14.3$\\
TXS 0828+193 & 127.7226 & 19.2210 & 2.572 & 10.1$\pm$1.0\tablenotemark{a} & 121.6$\pm$3.4 & 30$\pm$3\tablenotemark{b} & 91.6$\pm
$4.5 & $-2.9<\theta<6.1$\\
MRC 2025-218 & 306.9974 &-21.6825 & 2.63  & 8.3$\pm$2.3\tablenotemark{e} & 93.0$\pm$8.0 & 7$\pm$5\tablenotemark{b} & 86$\pm$9 &
$-13<\theta<5$\\
TXS 0943-242 & 146.3866 &-24.4804 & 2.923 & 6.6$\pm$0.9\tablenotemark{a} & 149.7$\pm$3.9 & 60$\pm$2\tablenotemark{b} & 89.7$\pm$
4.4 & $-4.7<\theta<4.1$\\
TXS 0119+130 & 20.4280 & 13.3494 & 3.516 & 7.0$\pm$1.0\tablenotemark{f} & 0$\pm$15 & 85$\pm$5\tablenotemark{g} & 95$\pm$16 & $-1
1<\theta<21$\\
TXS 1243+036 & 191.4098 &  3.3890 & 3.570 & 11.3$\pm$3.9\tablenotemark{a} & 38.0$\pm$8.3 & 132$\pm$3\tablenotemark{b} & 86.0$\pm
$8.8 & $-12.8<\theta<4.8$\\
\tableline
Mean & & &2.80& & & & 89.2$\pm$2.2 & $-3.0<\theta<1.4$\\

\enddata
\tablecomments{The last row shows the mean for all RG.}
\tablenotetext{a}{\citet{Ver00}}
\tablenotetext{b}{\citet{Pen98}}
\tablenotetext{c}{\citet{Cim98}}
\tablenotetext{d}{\citet{Kno97}}
\tablenotetext{e}{\citet{Cim93}}
\tablenotetext{f}{C. De Breuck (2009, private communication)}
\tablenotetext{g}{\citet{DeB02}}
\end{deluxetable}

\end{document}